# Computación distribuida aplicada a la identificación de nuevos medicamentos

# Distributed computing applied applied to the identification of new drugs

Raul Isea[1*] y Rafael Mayo[2]

(1) Instituto de Estudios Avanzados IDEA, Hoyo de la Puerta, Baruta, Venezuela

(2) CIEMAT, Av. Complutense 22, Madrid, España

Dirección de Correspondencia:

Raúl Isea

Instituto de Estudios Avanzados IDEA, Valle de Sartenejas,

Hoyo de la Puerta, Baruta, Venezuela

Email: risea@idea.gob.ve

**Resumen**

El presente trabajo destaca las ventajas de implementar una computación distribuida para el uso intensivo en cómputo científico dedicado a la identificación de posibles medicamentos que pueden ser aplicados a resolver problemas en salud pública.

Palabras Claves: Computación Grid, Diseño de Medicamentos, EELA, WISDOM, Computación Distribuida.

**Abstract**

This work emphasizes the assets of implementing the distributed computing for the intensive use in computational science devoted to the search of new medicines that could be applied in public healthy problems.

**Introducción**

En los últimos años se ha destacado en la literatura científica el uso intensivo de cálculos computacionales para la resolución de problemas científicos. Claros ejemplos se han presentado en el campo de la física de partículas (Adamová *et al*., 2010) o la deducción de historias evolutivas (Isea *et al*., 2009), así como la búsqueda automática de secuencias similares en genomas completos (Aparicio *et al*., 2009). Asimismo se pueden encontrar diversos trabajos dedicados a la identificación de nuevos medicamentos empleando para ello el computador como herramienta experimental, como se indicó en un estudio realizado por Jiang y colaboradores para la búsqueda de nuevos medicamentos efectivos contra el HIV-1 (Jiang *et al.*, 2006). Dicho cálculo fue realizado en el supercomputador Blue Gene gracias a una inversión realizada por IBM (considerado el más rápido en el mundo en el momento en que se realizó dicha investigación) y para llevarlo a cabo se emplearon 16.384 procesadores del tipo PowerPC 440 dual-core, empleando el programa DOCK (Moustakas *et al.*, 2006). Otro ejemplo fue el cálculo que se realizo en el supercomputador de San Diego para la identificación de nuevos medicamentos contra la gripe aviar (Cheng *et al.*, 2008). En este caso se identificaron 27 potenciales fármacos similares a Tamiflu con el programa AutoDOCK (Morris *et al.*, 1996) gracias al uso intensivo y dedicado de 128 procesadores del tipo PowerPC.

Estos dos ejemplos están caracterizados por la alta inversión requerida para la adquisición de dichas infraestructuras computacionales, sin pasar por alto los gastos necesarios para la formación de especialistas y profesionales dedicados a tiempo completo a la instalación, puesta a punto y administración de dichos recursos.

**Primeros pasos en la Computación Grid**

La Computación Grid se caracteriza por implementar un protocolo de trabajo donde el usuario dispone de una amplia gama de recursos computacionales sin que él conozca los requerimientos técnicos necesarios para la ejecución del mismo. Esta filosofía de trabajo comenzó con un planteamiento inicial propuesto por Leonard Kleinrock desde 1969, quien estuvo trabajando en el desarrollo del primer navegador Web llamado Mosaic donde él especulaba que llegaría un día en que los servicios de computación estarían prestando un servicio semejante a los servicios esenciales como la luz o el teléfono. Veintinueve años después, Ian Foster y Carl Kesselman propusieron el concepto de Computación Grid al definirlo como aquella infraestructura computacional que esta disponible a los usuarios que así lo requieran, independientemente de donde estén ubicados los mismos y que coordina recursos que no están bajo un control centralizado, usa estándares libres para protocolos e interfaces y proporciona calidades de servicio no triviales (Foster *et al.*, 1999; Foster, 2002). Por ejemplo, un usuario puede realizar un cómputo científico desde la República Bolivariana de Venezuela desconociendo el lugar físico donde se realizara el mismo. Esta última afirmación es posible gracias a iniciativas como EELA, abreviatura que corresponde a *E-infraestructura compartida entre Europa y Latinoamérica*, y sus secuelas, las cuales han sido financiadas por la Comisión Europea (Ciuffo *et al.*, 2009; Hernández *et al.*, 2007). No obstante, y a modo de ejemplo ilustrativo, mencionar que el monto económico para la realización de la segunda fase de EELA (conocida como EELA-2), desde abril de 2008 hasta marzo de 2010, ascendió a la suma total de 4.869.875 euros. Gracias a dicho financiamiento se implementó una infraestructura de cómputo científico en producción equivalente a 6.100 procesadores con una capacidad de 560 Petabytes de almacenamiento de datos.

**Primeros pasos en la Computación Distribuida**

Esta nueva filosofía en computación se está empleando en diversos problemas que requieran grandes demandas computacionales, y el ejemplo más conocido es el programa SETI@Home, al estar implementado como un salvapantalla sin que el usuario conozca los detalles técnicos del cálculo que se está realizando en dicho computador. En este sentido, recordemos que SETI@Home fue diseñado para la búsqueda de inteligencia extraterrestre con datos provenientes de radiotelescopios (Korpela *et al.*, 2004), donde se han registrado más de 4 millones de computadoras que han empleado dicho sistema de cómputo.

Gracias a esta filosofía de trabajo computacional, dicho proyecto está empleando un sistema de cómputo como sí fuera un solo "supercomputador virtual" compuesto por una gama de computadoras cuyo requisito es que estén conectadas en Internet. Esta idea novedosa es atribuida a David Gedye en 1995, aunque oficialmente se lanza el proyecto cuatro años después. En este sentido, es importante resaltar que la versión original del salvapantalla SETI@Home se mantuvo originalmente hasta el 2005. Posteriormente, y debido al éxito alcanzado, se implementaron nuevos middleware capaces de gestionar los trabajos de cómputo más eficiente, lo que derivo al proyecto "SETI@Home Enhanced" (más detalles disponible en la página Web ubicada en http:\\setiathome.berkeley.edu/sah_enhanced.php).

Posteriormente se implementaron otros programas a raíz del éxito de esta metodología tales como Folding@Home (detalles disponibles en Internet a través de la dirección Web http:\\folding.stanford.edu). Este proyecto comenzó en el año 2000 en la Universidad de Stanford bajo la supervisión del Profesor Vijay Pande y su objetivo es la simulación del plegamiento de péptidos y proteínas (Voelz *et al.*, 2010). Antes de impulsar este tipo de iniciativas computacionales, el mismo se realizaba en supercomputadoras. Citemos por ejemplo el cálculo realizado por Duan y colaboradores que simularon el plegamiento de un péptido de tan solo 37 aminoácidos, el cual fue realizado en una supercomputadora Cray donde se dedicaron 256 procesadores para ello, con un tiempo de cómputo de dos meses consecutivos (Duan *et al.*, 1998).

Otro tipo de computación distribuida es la computación voluntaria. En ella, se aprovecha el tiempo de inactividad de los ordenadores personales para que estos ejecuten tareas de interés para un proyecto común, el cual tiene sus servidores centralizados que procesan los distintos trabajos y almacenan los resultados. A resultas de esta filosofía de trabajo se desarrolló el middleware de código abierto BOINC (de sus siglas en Inglés Berkeley Open Infrastructure for Network Computing), desarrollado por un grupo de la Universidad de California (Berkley) dirigido por David Anderson. Más detalles sobre esta plataforma se pueden encontrar en http://boinc.berkeley.edu/.

**Computación Distribuida *versus* Computación Grid**

Es importante aclarar dos conceptos que se están manejando indistintamente y los mismos no son equivalentes entre sí, es decir: Computación Grid y Computación Distribuida. Recordemos que la Computación Distribuida se refiere principalmente a la infraestructura donde se manejan varios sistemas de infraestructura computacional conectados por una red que se administran o no de manera individual (Naor *et al.*, 1995); mientras que la Computación Grid posee la filosofía que los recursos computacionales se comparten y administran independientemente de la zona geográfica donde se ubiquen los equipos (Foster *et al.*, 1999). Por ello, la Computación Grid necesita un sistema de gestión de trabajos para garantizar el uso óptimo de los recursos el cual dependerá de los requerimientos del usuario, conocido como middleware (Talia *et al.*, 2008).

Otro ejemplo erróneo en la conceptualización de la Computación Grid es aquella donde se concibe un único lugar físico con dos o más sistemas de cómputo científico que están interconectadas entre sí. Dicha arquitectura es un claro ejemplo de Computación Distribuida, porque se centraliza en un mismo lugar físico los equipos y porque normalmente se administran de forma única. Más aún, la mera instalación de programas especialmente diseñados para el manejo de aplicaciones a través del Grid, como por ejemplo, las herramientas de Globus Toolkit (Mache, 2006) o gestores de colas de trabajo tales como Condor (detalles disponibles en http://www.cs.wisc.edu/condor/) o Sun Grid Engine (disponible en http://gridengine.sunsource.net), tampoco aseguran la condición de Computación Grid. De hecho, sí los cálculos se realizan en una misma ubicación física, en este caso, el mismo debería enmarcarse dentro del concepto de Computación Ubicua, en el sentido de integrar la información sin distinción del equipo que se maneje, pero centralizado su administración (Greenfield, 2006).

La computación ubicua (empleado en lugar de la dupla en inglés *pervasive computing*) comparte una visión de equipos pequeños y baratos conectados de forma robusta, distribuidos a distinta escala para la vida diaria o no, pero siempre situados en puntos finales. Así por ejemplo, y a modo ilustrativo, una computación ubicua doméstica se referiría a la interconexión entre los interruptores de la luz o la calefacción, y los controles de las condiciones climatológicas presentes para adecuar la intensidad de los primeros a las necesidades particulares en una habitación. Otro escenario posible podría ser el de neveras que estuvieran censando los alimentos que están enfriando para así recomendar menús dependiendo de la condición del estado de conservación de dichos alimentos.

**Computación Grid aplicada en Malaria**.

La malaria o paludismo es la responsable de la muerte de entre dos a tres millones de personas anualmente, donde un alto porcentaje corresponde al fallecimiento de niños menores de cinco años. Por ello se ha dedicado un esfuerzo para la búsqueda de nuevos medicamentos gracias a la aplicación de la tecnología Grid. En este sentido, se debe resaltar la iniciativa internacional denominada WISDOM (abreviatura del inglés que significa *Wide In-Silico Docking Of Malaria*) para la identificación de posibles medicamentos útiles contra la malaria. Dicha iniciativa emplea el programa AutoDOCK (Morris et al 1996), y hasta la fecha, ya se han celebrado dos ensayos computacionales a gran escala (conocidos como *data challenges*) para distintas dianas en malaria (detalles en la página Web http://wisdom.healthgrid.org).

En su segundo ensayo conocido como WISDOM-II, se dispuso de un cálculo intensivo donde se ensayaron 4.3 millones de drogas conocidas por cada proteína blanco implicada en la recepción de posibles medicamentos (detalles en el trabajo de Kasam *et al.*, 2009). El mismo se terminó realizando en aproximadamente 5.000 procesadores que estaban distribuidos en 17 países. A pesar de disponer de varios miles de procesadores, finalizó en 76 días consecutivos. Sí se hubiera replicado dicho cálculo en un solo computador, el tiempo estimado *grosso modo* hubiera sido de 413 años (Salzemann *et al.*, 2007). Se debe destacar la rapidez en la ejecución de dicho ensayo computacional gracias al esfuerzo de diversos equipos transdisciplinarios de diversos países europeos, africanos y latinoamericanos.

No obstante, se presenta un nuevo desafió por el volumen de información que se está generando debido a la implementación de dicha metodología, porque en el caso de WISDOM-II, por citar un ejemplo, se requiere analizar un poco más de 1.700 Gigabytes de datos. En ese sentido se están empleando metadatos que permiten profundizar la calidad de los resultados en forma automática, como por ejemplo, el sistema llamado AMGA (Koblitz *et al*., 2007)

 **Conclusiones**

La computación Grid ha permitido innovar en muchos campos científicos, y un claro ejemplo de ello es la identificación de posibles medicamentos, gracias a que dicha tecnología permite unificar una serie de recursos computacionales que están distribuidos a lo largo de todo el mundo. Esta metodología se caracteriza por los bajos costos económicos en comparación con los gastos implicados en la adquisición y uso de supercomputadores. Más aún, dichas simulaciones permiten reutilizar ciertos diseños de medicamentos que puedan ser empleados en otras enfermedades no contempladas en su diseño original. Finalmente, destacar una limitación de uso de los programas implementados en Grid debido a que los mismos se deben diseñar, compilar, y configurar para dicho fin, y lamentablemente muchos programas actuales se han conceptualizado bajo un esquema de cómputo y administración centralizado.